\documentclass[prb,showpacs,preprintnumbers,amssymb,twocolumn]{revtex4}
\usepackage{graphicx}% Include figure files
\usepackage{dcolumn}% Align table columns on decimal point
\usepackage{bm}% bold math

\usepackage{amsfonts}

\def\half{\textstyle{\frac{1}{2}}}
\def\quarter{\textstyle{\frac{1}{4}}}

\def\H{{\cal H}}

\def\p{\phi}

\def\H{{\cal H}}

\def\v{\vskip.3cm}

\def\l{\lambda}

\def\t{\textstyle}

\def\ra{\rightarrow}
\def\tint{{\textstyle\int}}

\def\s{\hskip.08em}

\def\d{\partial}

\def\b{\begin{eqnarray*}}  %takes no eqn numbers
\def\e{\end{eqnarray*}}    %takes no eqn numbers
\def\bn{\begin{eqnarray}}  %takes eqn numbers
\def\en{\end{eqnarray}}   %takes eqn numbers
\def\<{\langle}
\def\>{\rangle}

\def\no{\nonumber}

\def\k{\kappa}
\def\{{\lbrace}
\def\}{\rbrace}
\begin{document}
\title{Proposal for Divergence-Free \\Quantization of Covariant Scalar Fields}
\author{John R. Klauder}
\affiliation{Department of Physics and Department of Mathematics,
University of Florida,
Gainesville, FL 32611-8440}
\date{ }
%let\frak\cal

%\renewcommand{thesection}{\Roman{section}.}
%\renewcommand{thesubsection}{\Alph{subsection}.}
%\renewcommand{\baselinestretch{2}}        %for double spacing
\bibliographystyle{unsrt}
%\begin{document}
%\maketitle
\begin{abstract}
Guided by idealized but soluble nonrenormalizable models, a nontraditional proposal for the quantization of
covariant scalar
field theories is advanced, which achieves a term-by-term, divergence-free perturbation analysis of interacting
models expanded about a suitable pseudofree theory [differing from a free theory by an $O(\hbar^2)$ term]. This procedure not only provides acceptable solutions for models for which no acceptable solution
currently exists, e.g., $\varphi^4_n$,
for spacetime dimension $n\ge4$, but offers a new, divergence-free solution, for less-singular models as well,
e.g., $\varphi^4_n$, for $n=2,3$.
\end{abstract}

\maketitle
%\section{Introduction}
It is common knowledge that divergences arise in the study of covariant quantum field theories, and elaborate
efforts (e.g., supersymmetry) are used to nullify the effects of these divergences.  In this letter we argue that adopting an
appropriate  $\hbar$ ambiguity in the quantization procedure can eliminate the divergences that are usually
encountered. Although we focus on scalar fields, similar methods may apply for other
quantum field theories. As motivation for our approach, we initially analyze how divergences are eliminated
in soluble ultralocal models.\v
{\it Ultralocal Models}\v     %\section*{Ultralocal Models}
The classical action for the quartic ultralocal model is given by
  \bn  A=\int\{\s\half[{\dot\phi}(t,{\bf x})^2-m_0^2\s\phi(t,{\bf x})^2]
  -g_0\s\phi(t,{\bf x})^4\s\}\,dt\s d{\bf x}\;,\en
  where $g_0\ge0$, ${\dot\phi}(t,{\bf x})=\d\phi(t,{\bf x})/\d t$, ${\bf x}\in{\mathbb R}^s$, and
$1\le s<\infty$. With no spatial gradients, the light cone
of covariant models collapses to a temporal line reflecting the statistical independence of ultralocal fields at
any two distinct spatial points. This vast symmetry ultimately helps determine the quantum theory for such models.

Viewed conventionally, it is hard to imagine a quartic interacting field theory that would cause more trouble
in its quantization. On one hand,
 it is clear that ultralocal models are perturbatively nonrenormalizable for any $s\ge1$; on the other
hand, if viewed nonperturbatively, and limited to mass and coupling constant renormalizations, they lead to
free (Gaussian) results based simply on the Central Limit Theorem. Clearly, other methods are required.

Although the quantum theory of these models has been completely solved without introducing cutoffs \cite{ul,book},
it is pedagogically useful to study the model as regularized by a hypercubic spacetime lattice with periodic
boundary conditions. If $a>0$ denotes the lattice spacing and $L<\infty$ denotes the number of sites on each
edge, then the ground-state
distribution of the free theory ($g_0\equiv0$) is described by the characteristic function
\bn   C_f(f)\hskip-1em&&=M\int e^{ i\Sigma'_k f_k\s \phi_k\s a^s-m_0\Sigma'_k \phi_k^2\s a^s}\,\Pi'_kd\phi_k
\no\\
     &&=e^{ -(1/4\s m_0)\Sigma'_k f_k^2\s a^s}\no\\
     &&\ra e^{-(1/4\s m_0)\s\int f({\bf x})^2\,d{\bf x}}
\;, \en   where in the last line we have taken the continuum limit.
In this expression $k=(k_1,k_2,\ldots,k_s)$, $k_j\in{\mathbb Z}$, labels the sites in this spatial lattice.
It is of interest to calculate mass-like moments in the ground-state distribution as given by
  \bn I_p(m_0)\hskip-1em&&\equiv M\int [\Sigma'_k\phi_k^2\s a^s]^p\,e^{ -m_0\Sigma'_k \phi_k^2\s a^s}\,\Pi'_kd\phi_k\no\\
     &&=O((N'/m_0)^p)\,I_0(m_0)\;, \label{emoments}\en
where $N'\equiv L^s$ is the number of lattice sites in the spatial volume. A perturbation of the mass, with
$\Delta\equiv {\tilde m}_0-m_0$, leads to
  \bn  I_1({\tilde m}_0)=I_1(m_0)-\Delta\s I_2(m_0)
             +\half\s\Delta^2\s I_3(m_0)-\cdots\;, \en
which, assuming $m_0=O(1)$,  exhibits increasingly divergent contributions in the continuum limit in which
$a\ra0$ and $L\ra\infty$
such that $N'\s a^s=(La)^s$ remains large but finite. The origin of these divergences is exposed if we pass
to hyperspherical coordinates, where $\phi_k\equiv\k\s \eta_k$, $\Sigma'_k\phi_k^2\equiv\k^2$, and
$\Sigma'_k\eta_k^2\equiv1$, for which (\ref{emoments}) becomes
   \bn &&\hskip-2em I_p(m_0)=2\s M\int \k^{2p}\s a^{sp}\,e^{ -m_0\s\k^2\s a^s}\,\k^{(N'-1)}\s d\k\no\\
   &&\hskip6em\times\delta(1-\Sigma'_k
\eta_k^2)\,\Pi'_kd\eta_k\;, \label{ekappa}\en
which not only reveals the source of the divergences as the term $N'$ in the measure factor $\k^{(N'-1)}$,
but also confirms the approximate evaluation of (\ref{emoments}) by a steepest descent analysis of the $\k$
integration. If we could somehow change the power of $\k$ in the measure of (\ref{ekappa}) to $\k^{(R-1)}$,
where $R$ is a finite factor, these divergences would be eliminated!

The theory of infinite divisibility \cite{def} assures us that besides the Gaussian ground-state distributions
there are only Poisson ground-state distributions that respect the ultralocal symmetry of the model, and they are
described by characteristic functions of the form
  \bn  C(f)=e^{-\tint d{\bf x}\s\tint[\s1-\cos(f({\bf x})\s\l)\s]\,c(\l)^2\,d\l}\;, \en
where $\tint[\l^2/(1+\l^2)]\,c(\l)^2\,d\l<\infty$, but $\tint c(\l)^2\,d\l=\infty$ (to ensure the smeared field
operator only has a continuous spectrum). As an
important example, let us assume that $c(\l)^2=\half b\s\exp(-b\s m\s\l^2)/|\l|$,
where $b$ is a positive constant with dimensions (Length)$^{-s}$, and $m$ is a mass parameter. For this example,
it follows that
\bn &&M'\int e^{ i\Sigma'_k f_k\phi_k\s a^s -m_0\Sigma'_k\phi_k^2\s a^s}
   \Pi'_k[\s|\phi|^{(1-ba^s)}\s]^{-1}\,\Pi'_kd\phi_k \no \\
    &&\hskip0em =\Pi'_k \{1-(\half b\s a^s)\tint[\s1-\cos(f_k\s\l)\s]
    \, e^{ -b m\l^2}\,d\l/|\l|^{(1-ba^s)}\}\no\\
   &&\hskip0em \ra e^{-\half b\tint d{\bf x}\s\tint[\s1-\cos(f({\bf x})\s\l)\s]\,
e^{ -bm\l^2}\,d\l/|\l|}\;;\label{emean}\en
here we have set $m_0=ba^s\s m$, $\l=\phi\s\s a^s$, and used the fact that to leading order $M'=(\half ba^s)^{N'}$,
which holds because
   \bn (\half ba^s)\tint e^{-bm\l^2}\,d\l/|\l|^{(1-ba^s)}
   \hskip-1em&&\simeq (ba^s)\tint_0^B d\l/\l^{(1-ba^s)}\no\\
   && =B^{ba^s}\ra1\;,\en
provided that $0<B<\infty$.

Observe that the lattice ground-state distribution for this example is
\bn \frac{(\half b a^s)^{N'}\,e^{-m_0\Sigma'_k\phi_k^2\s a^s}}{\Pi'_k|\phi_k|^{(1-ba^s)}}
  =\frac{(\half b a^s)^{N'}\,e^{-m_0\k^2\s a^s}}{\k^{(N'-ba^s\s N')}\,\Pi'_k|\eta_k|^{(1-ba^s)}}
\label{egsdist}\en
which has exactly the right factor to change the $\k$ measure from $\k^{(N'-1)}$ to $\k^{(R-1)}$, where
in the present case $R=ba^s N'$ [a finite number chosen in order to ensure a meaningful continuum limit
for (\ref{emean})].  If we adopt (\ref{egsdist}) as the appropriate ``pseudofree''
ground-state distribution, then all divergences due to integration over $\k$ will disappear!\v

{\it Free and Pseudofree Theories}\v    % \subsubsection*{Free and Pseudofree Theories}
What exactly do we mean by free and pseudofree theories?
An elementary example of a theory that involves pseudofree behavior is given by
the anharmonic oscillator with the classical action
   \bn A=\tint\{\half[{\dot x}(t)^2-x(t)^2]-g_0\s x(t)^{-4}\}\,dt\;,  \en
where $g_0\ge0$.
The free theory ($g_0\equiv0$) has solutions $A\cos(t+\gamma)$ that freely cross $x=0$; when $g_0>0$, however, {\it no}
solution can cross $x=0$, and the limit of the interacting solutions as $g_0\ra0$ becomes
$\pm\s |A\cos(t+\gamma)\s|$.
This latter behavior describes the classical pseudofree model, i.e., the model continuously connected to
the interacting models as $g_0\ra0$.
Quantum mechanically, the imaginary-time propagator for the free theory is given by
  \bn  K_f(x'',T;x',0)
  ={\t\sum}_{n=0}^\infty h_n(x'')\s h_n(x')\,e^{-(n+1/2)T}\;, \en
where $h_n(x)$ denotes the $n$th Hermite function. However, for the interacting quantum theories,
as the coupling $g_0\ra0$, the imaginary-time propagator converges to
  \bn &&\hskip-4em K_{pf}(x'',T;x',0)=\theta(x''x')\,{\t\sum}_{n=0}^\infty \s h_n(x'')\no\\
  &&\hskip+1em \times [h_n(x')-h_n(-x')\s]\,e^{-(n+1/2)T}\;,\en
where $\theta(y)=1$ if $y>0$ and $\theta(y)=0$ if $y<0$,
which characterizes the quantum pseudofree model. This behavior has arisen because within a functional integral
the interaction acts
partially as a hard core projecting out certain histories that would otherwise be allowed by the free theory; any
perturbation analysis of the interaction term clearly must take place about the pseudofree theory and not
about the free theory. The field theory models are more involved, but the basic ideas are essentially the same.\v

{\it Lessons from Ultralocal Models} \v  %\subsubsection*{Lessons from Ultralocal Models}
Observe for the classical ultralocal models that when $g_0>0$ it is necessary that
$\tint\phi(t,{\bf x})^4 dt\s d{\bf x}<\infty$
to derive the equations of motion, but when $g_0=0$ this restriction is absent. Thus the set of classical
solutions for $g_0>0$ does {\it not} reduce as $g_0\ra0$ to the set of classical solutions of the free theory;
instead, the set of classical solutions for $g_0>0$ passes by continuity to a set of classical solutions of the
free theory that also incorporates the hard-core consequences of the condition
$\tint\phi(t,{\bf x})^4 dt\s d{\bf x}<\infty$. An interacting classical theory that is not
continuously connected to its own free classical theory is likely to be associated with
an interacting quantum theory that is not continuously connected to its own free quantum theory. This situation is
easy to see for the ultralocal models. The
characteristic function of the ground-state distribution has either a Gaussian or a Poisson form as indicated,
and there is no continuous, reversible path between the two varieties. If one seeks nontriviality, then the
interacting theory must be of the Poisson type; and as the coupling constant vanishes, the continuous limit must
also be a Poisson distribution, namely, the pseudofree model as chacterized by (\ref{emean}).

To complete the ultralocal story, we observe that the ground-state distribution for interacting models is also of
the Poisson form, where
\bn c(\l)^2=\half b\,\exp[-y(\l)]/|\l|\en
 for suitable functions $y(\l)$. Each such distribution leads to a lattice Hamiltonian and thereby a lattice
action for a full (Euclidean) lattice spacetime functional integral formulation. The pseudofree model has the
lattice action of a traditional free theory augmented by a local
counterterm proportional to $\hbar^2$ and (surprise!) {\it inversely} proportional to the field squared, so
that it accounts for the denominator factor in (\ref{egsdist}), which is central to an overall divergence-free
formulation. The form of the nontraditional counterterm is implicitly given in the next section, and since these
models have been extensively discussed elsewhere \cite{ul,book}, we do not pursue them further.

However, we do take from the ultralocal model {\it the central principle of our analysis}, which we
dub ``measure mashing''.
In particular, in extending our analysis to covariant models, we adopt the ``slick trick''
that worked so well
for the ultralocal models, namely, choosing a pseudofree model that changes the measure factor $\k^{(N'-1)}$
for the hyperspherical radius to the form $\k^{(R-1)}$, where $R$ is a suitable finite factor for the model in
question.\v

{\it Covariant Models} \v  %\subsubsection*{Covariant Models}
We restrict our initial attention to models with the classical
              action given by
\bn &&\hskip-1em A=\tint(\half\s\{{\dot\phi}(x)^2-[{{\nabla}\phi}(x)]^2-m^2_0\s\phi(x)^2\s\}\no\\
               &&\hskip5.74em -\l_0\s\phi(x)^4\s)
\,d^n\!x\;,\en
               where $x=(t=x_0,x_1,x_2,\ldots,x_s)\in{\mathbb R}^n$, $n=s+1\ge5$, $\l_0\ge0$, ${\dot\phi}(x)=\d\phi(x)/\d t$,
and $[{\nabla\phi}(x)]^2\equiv\Sigma_{j=1}^s
               (\d\phi(x)/\d x_j)^2$. It is not obvious, but for the spacetime dimensions in question, the
interaction term imposes a restriction on the free action as follows from
               the multiplicative inequality \cite{lady,book}
               \bn &&\hskip-1em\{\tint \phi(x)^4\,d^n\!x\}^{1/2}\no\\
               &&\hskip2em\le C\tint\{{\dot\phi}(x)^2+[\nabla\phi(x)]^2+\phi(x)^2\s\}
\,d^n\!x\;, \en
               where for $n\le4$ (the renormalizable cases), $C=4/3$ is satisfactory, while for
               $n\ge5$ (the nonrenormalizable cases), $C=\infty$ meaning that there are fields for which the
left side diverges while the right side is finite (e.g., $\phi_{singular}(x)=|x|^{-p}\,e^{-x^2}$,
               where $n/4\le p< n/2-1$). As a consequence, for $n\ge5$ the set of solutions of the interacting
classical theory does {\it not} reduce to the set of solutions of the free classical theory as the coupling
constant $g_0\ra0$. We now examine the quantum theory in the light of this knowledge, and we initially focus
on finding a suitable pseudofree model for covariant theories.\v

{\it Choosing the Covariant Pseudofree Model}\v  % \subsubsection*{Choosing the Covariant Pseudofree Model}
              For covariant scalar fields, the lattice version of a free, nearly massless, quantum theory has a
              characteristic functional for the ground-state distribution given by
              \bn  C_f(f)
              =M'\int e^{ i\Sigma'_kf_k\phi_k\s a^s -\Sigma'_{k,l}\phi_k\s A_{k-l}\s\phi_l\s\s
a^{2s}}\,\Pi'_kd\phi_k\:,\en
              where $A_{k-l}$ accounts for the derivatives and a small, well-chosen, artificial mass-like
contribution. The quantum Hamiltonian for this ground state
              (restoring $\hbar$) becomes
              \bn \hskip-.6em\H_f=-\half\hbar^2 a^{-s}{\t\sum'_k}\s\frac{\t\d^2}{\t\d\phi_k^2}
              +\half{\t\sum'_{k,l}}
\phi_k\s A^2_{k-l}\s\phi_l\s\s a^{3s}-E_0 \en
              where $E_0$ is a constant ground state energy and
              \bn  A^2_{k-l}\hskip-1em&&\equiv \Sigma'_p A_{k-p}\s A_{p-l}\no\\
              &&\hskip0em\equiv\Sigma_{j=1}^s[\s 2\s\delta_{k,l}
-\delta_{k+\delta_j,l}-
              \delta_{k-\delta_j,l}\s]\s a^{-(2s+2)}\no\\
              &&\hskip5em + s\s L^{-2s}\s\delta_{k,l}\s a^{-(2s+2)}\;,\en
              where $k\pm\delta_j\equiv (k_1,k_2,\ldots,k_j\pm1,\ldots,k_s)$, and the last factor is
a small, artificial mass-like term (introduced to deal with the zero mode $\p_k\ra\p_k+\s\xi$). The true mass
term will be introduced later along with the quartic interaction when we discuss the final model.

              We next modify
              the free ground-state distribution in order to suggest a suitable characteristic function
for the pseudofree ground-state
              distribution by the expression
              \bn &&\hskip-1em C_{pf}(f)
              =M''\int e^{ i\Sigma'_kf_k\phi_k\s a^s -\Sigma'_{k,l}\phi_k\s A_{k-l}\s\phi_l\s
\s a^{2s}}\no\\  &&\hskip7
em\times  e^{ -W(\phi\s\s a^{(s-1)/2}/\hbar^{1/2})}\,\no\\
              &&\hskip1em\times\{\Pi'_k[\Sigma'_l J_{k,l}\s\phi_l^2]\s\}^{-(1-R/N')/2}\,\Pi'_kd\phi_k\;,
\label{ecovpf}\no\\
\en
              where the constants $J_{k,l}\equiv 1/(2s+1)$ for the $(2s+1)$ points that include $l=k$ and all
              the $2s$ spatially nearest neighbors to $k\s$; $J_{k,l}\equiv 0$ for all other points. Stated
otherwise,
              the term $\Sigma'_l J_{k,l}\s\phi^2_l$ is {\it an average  of field-squared values} at $l=k$ and
the $2s$ spatially nearest neighbors to $k$. Note well, that this term leads to a factor of $\k^{-(N'-R)}$ that,
in effect,
              replaces the hyperspherical radius variable measure term $\k^{(N'-1)}$ by the factor $\k^{(R-1)}$
(i.e., mashing the measure), and since $R$ is finite, this choice
              eliminates any divergences caused by integrations over the variable $\k$. Guided by the
              ultralocal models,
              we choose the finite factor $R=b a^sN'$ in an initial effort to find suitable pseudofree models
              for the covariant theories. The factor $A_{k-l}$ is the same as introduced for the
              free theory, while the function $W$ is implicitly defined below.\v

{\it The Hamiltonian for the Covariant Pseudofree Model}\v
%\subsubsection*{The Hamiltonian for the Covariant Pseudofree Model}
              The Hamiltonian follows from the proposed ground state wave function contained in (\ref{ecovpf}).
 To understand the role
              played by $W$, let us first assume that $W=0$. Then, in taking the necessary second-order
derivatives, there will be a contribution when one derivative acts on the
              $A_{k-l}$ factor in the exponent and the other derivative acts on the denominator factor
              involving $J_{k,l}$. The result will be a cross term that exhibits a long-range interaction
              that would cause difficulty for causality in the continuum limit. Instead, at this point,
              we focus on the Hamiltonian itself as primary (rather than the ground state), and adopt the
              Hamiltonian for the pseudofree model as
               \bn &&\hskip-1em\H_{pf}=-\half\hbar^2\s a^{-s}\s{{\t\sum}}'_k\frac{\t\d^2}{\t\d\phi_k^2}
              +\half\s{\t\sum'}_k(\phi_{k^*}-\phi_k)^2\s a^{s-2}\no\\
              &&\hskip1em+\half s(L^{-2s}\s a^{-2}){\t\sum'}_k\phi_k^2\s a^s
               +\half\s \hbar^2{\t\sum'}_k{\cal F}_k(\phi)\s a^s-E_{pf}\;,\no\\
                   && \en
                 where $k^*$ represents a spatially nearest neighbor to $k$ in the positive sense, implicitly
summed over all $s$ spatial directions, and the counterterm ${\cal F}_k(\p)$, which follows from both
                 derivatives acting on the $J_{k,l}$ factor, is given by
                 \bn  &&\hskip-1em{\cal F}_k(\p)
\equiv{\quarter}\s(1-ba^s)^2\s
          a^{-2s}\s\bigg({\t\sum'_{\s t}}\s\frac{
  J_{t,\s k}\s \p_k}{[\Sigma'_m\s
  J_{t,\s m}\s\p_m^2]}\bigg)^2\no\\
  &&\hskip2em-{\half}\s(1-ba^s)
  \s a^{-2s}\s{\t\sum'_{\s t}}\s\frac{J_{t,\s k}}{[\Sigma'_m\s
  J_{t,\s m}\s\p^2_m]} \no\\
  &&\hskip2em+(1-ba^s)
  \s a^{-2s}\s{\t\sum'_{\s t}}\s\frac{J_{t,\s k}^2\s\p_k^2}{[\Sigma'_m\s
  J_{t,\s m}\s\p^2_m]^2}\;. \label{eF} \en
  We observe that this form for the counterterm leads to a local potential in the continuum limit
  even though it is a rather unfamiliar one. ({\bf Remark:} If $J_{k,l}$ is taken as $\delta_{k,l}$, the resultant counterterm is that appropriate to the ultralocal models.)

  With this involved counterterm, the pseudofree Hamiltonian is completely defined, and we define the implicitly
given expression for the pseudofree ground state to be the ground state $\Psi_{pf}(\p)$ for this Hamiltonian. For large
  $\phi$ values the $A_{k-l}$ term well represents the solution, and for small $\phi$ values the denominator
term involving the  $J_{k,l}$ term also well represents the solution. The role of the (unknown) function $W$
and $E_{pf}$ is to fine tune the solution so that it satisfies the equation $\H_{pf}\s \Psi_{pf}(\p)=0$. The
manner in which both $a$ and $\hbar$ appear in $\H_{pf}$ dictates how they appear in
  $W$ as $W(\phi\s\s a^{(s-1)/2}/\hbar^{1/2})$.\v
{\it Final Form of Lattice Hamiltonian and Lattice Action}\v
  %\subsubsection*{Final Form of Lattice Hamiltonian and Lattice Action}
  It is but a small step to propose expressions for the lattice Hamiltonian and lattice action in
  the presence of the proper mass term and the quartic interaction. The lattice Hamiltonian is given by
  \bn &&\hskip-1em\H=-\half\hbar^2\s a^{-s}\s{{\t\sum}}'_k\frac{\t\d^2}{\t\d\phi_k^2}
               +\half\s{\t\sum'}_k(\phi_{k^*}-\phi_k)^2\s a^{s-2}\no\\
               &&\hskip1.4em +\half s(L^{-2s}\s a^{-2}){\t\sum'}_k\phi_k^2\s a^s
                +\half m^2_0{\t\sum}'_k\phi^2_k\s a^s\no\\&&\hskip1.4em +\l_0\s{\t\sum}'_k\phi^4_k\s a^s
               +\half\s \hbar^2{\t\sum'}_k{\cal F}_k(\phi)\s a^s-E \;, \en
  and the Euclidean lattice action reads
                   \bn &&\hskip-.8em I(\p,a,\hbar)\no\\
               &&=\half\s{\t\sum}_k\s{\t\sum}_{k^*}(\phi_{k^*}-\phi_k)^2\s a^{n-2}
               +\half s(L^{-2s}\s a^{-2})
{\t\sum}_k\phi_k^2\s a^n\no\\
               &&\hskip.2em +\half m^2_0{\t\sum}_k\phi^2_k\s a^n+\l_0\s{\sum}_k\phi^4_k\s a^n
               +\half\s \hbar^2{\t\sum}_k{\cal F}_k(\phi)\s a^n, \label{eaction}\en
                   where the last sum on $k^{*}$, here made explicit, is a sum over all $n$ lattice directions
                   in a positive sense from the site $k$, and
in both expressions the counterterm ${\cal F}_k(\phi)$ is given in (\ref{eF}). When one studies the
full action, as in a Monte Carlo analysis, then the small, artificial mass-like term can be omitted.

The generating function for Euclidean lattice spacetime averages is given, as usual, by
   \bn &&\hskip-1em\< e^{ Z^{-1/2}\Sigma_k h_k\s\p_k\s a^n/\hbar}\>\no\\
   &&\hskip0em\equiv {\widetilde{M}}\int
e^{ Z^{-1/2}\Sigma_k h_k\s\p_k\s a^n/\hbar-I(\p,a,\hbar)/\hbar}\,\Pi_kd\phi_k \;, \label{elattice}
 \en
where $Z$ is the field strength renormalization constant and $\Sigma/\Pi$ (without primes) denotes a
sum/product over the full spacetime lattice now with $k=(k_0,k_1,k_2,\ldots,k_s)$,
where $k_0$ denotes the imaginary-time direction.
 Elsewhere \cite{kla4, klaFI}, we have studied the perturbation
analysis of (\ref{elattice}) and have determined that: (i) the proper {\it field strength
renormalization} is given by $Z=N'^{-2}(qa)^{1-s}$, (ii) the proper {\it mass renormalization} is given by
  $m_0^2=N'(qa)^{-1}\s m^2 $, and (iii) the proper {\it coupling constant renormalization} is given by
$\l_0=N'^3(qa)^{s-2}\s \l$.
  Here, $q$ denotes a positive constant with dimensions (Length)$^{-1}$, and $m$ and $\l$ represent
  finite physical factors. It is noteworthy that $Z\s m_0^2=m^2/[N'(qa)^s]$
and $Z^2\s \l_0=\l/[N'(qa)^s]$. The characterization of the model is now complete. ({\bf Remark:} Although
we have confined attention to models with quartic interactions, measure mashing also enables higher powers,
e.g., $\varphi^{44}_n$, $\varphi^{444}_n$, etc., to be handled just as well \cite{kla4}.)

Much has changed by passing from a free model to a pseudofree model as the center of focus. Traditionally, when
forming local products from free-field operators, normal ordering is used. On the contrary,
after measure mashing, the pseudofree field operators satisfy multiplicative renormalization, and no normal
ordering is involved. Indeed, the very coefficients $m_0^2$ and $g_0$ act partially as multiplicative
renormalization factors for the associated products involved. To say that there are no divergences means,
for example, that the expression $m_0^2\Sigma'_k\p_k^2\s a^s$ is well defined, and this fact is
established by ensuring that $m_0^2\s\Sigma'_k\<\p_k^2\>\s a^s\propto N'a^s<\infty$. The same holds true
for $\l_0\Sigma'_k \p_k^4\s a^s$, which is shown to be well defined by noting that
$\l_0\Sigma'_k \<\p_k^4\>\s a^s\propto N'a^s<\infty$. These quantities remain bounded even in the
continuum limit.\v
{\it Extension to Less Singular Scalar Models}\v
%\subsubsection*{Extension to Less Singular Scalar Models}
Let us take up the extension of measure mashing to other models such as $\varphi^4_n$, for  $n\le4$.
Although the classical pseudofree theory is identical to the classical free theory in these
cases, this fact does not prevent us from suggesting the consideration of mashing the measure for
such less singular models in an effort to eliminate divergences that arise in those cases. For $n=2$,
it is well known that normal ordering removes all divergences, but it is also well known that normal
ordering is a rather strange rule to define local products. In particular, if we rewrite the
product of two free-field operators as
\bn \varphi(x)\s\varphi(y)
=\<0|\s\varphi(x)\s\varphi(y)\s|0\>+:\varphi(x)\s\varphi(y): \;,\en
then, as $y\ra x$, the most singular term is the first term, but since it is a multiple of unity, the {\it second
and less singular term} is chosen to define the local product, $ \varphi(x)^2_{Renormalized}=\;\,:\varphi(x)^2:$; moreover, this expression is {\it not} positive despite being the chosen local ``square'' of the field.
In sharp contrast, in the operator  product expansion, schematically given by
  \bn &&\hskip-3.4em\varphi(x)\s\varphi(y)= c_1(x,y)\,\zeta_1(\half(x+y))\no\\
  &&\hskip1.2em+c_2(x,y)\,\zeta_2(\half(x+y))+\cdots\;, \en
the local product, as $y\ra x$, is defined as that local operator, say
$\varphi(x)^2_{Renormalized}=\zeta_1(x)$, for which the
associated $c$-number
coefficient $c_1(x,y)$ is the {\it most singular} as $y\ra x$; this is a very reasonable rule, and this choice is also positive as a square should be. To adopt measure mashing for $\phi^4_2$ would
introduce the operator product expansion and thereby a more natural local product definition. This same feature
also applies to $\p^4_3$ and $\p^4_4$, and moreover it would eliminate divergences that appear in all these
models. It could even offer a nontrivial proposal for $\p^4_4$ which is widely believed to be trivial
when quantized conventionally.

Is all this a physically realistic proposal? Presumably, the answer would depend on the application, so it
is too soon to expect a firm answer to this question. Nevertheless,
it would seem there is progress already just to have a possible solution to nonrenormalizable models
rather than the unsatisfactory results obtained by conventional techniques.

\end{document}